\documentclass[useAMS,usenatbib]{mn2e}
\usepackage{graphicx}
\usepackage{txfonts}
\usepackage{natbib}
\usepackage{wrapfig}
\usepackage{longtable}
\usepackage[usenames]{color}
\def\aj{AJ}%
\def\araa{ARA\&A}%
\def\apj{ApJ}%
\def\apjs{ApJS}%
\def\apss{Ap\&SS}%
\def\aap{A\&A}%
\def\aaps{A\&AS}%
\def\mnras{MNRAS}%
\def\prd{Phys.~Rev.~D}%
\newcommand{\bc}{\begin{center}}
\newcommand{\ec}{\end{center}}

\begin{document}
\title[The UV-optical colours of brightest cluster galaxies in optically and X-ray selected clusters]
      {The UV-optical colours of brightest cluster galaxies in optically and X-ray selected clusters}
\author[Jing ~Wang et al.]
       {Jing Wang$^{1,2}$\thanks{Email: wangj@mpa-garching.mpg.de},
       Roderik Overzier$^2$, Guinevere Kauffmann$^2$,
       \newauthor Anja von der Linden$^3$, Xu Kong$^{1,4}$     \\
       $^1$Center for Astrophysics, University of Science and Technology of China,
        230026 Hefei, China
       \\
       $^2$Max--Planck--Institut f\"ur Astrophysik,
        Karl--Schwarzschild--Str. 1, D-85748 Garching, Germany
        \\
       $^3$Kavli Institute for Particle Astrophysics and Cosmology,
   Stanford University, 452 Lomita Mall, Stanford, CA  94305-4085, USA
        \\
       $^4$Key Laboratory for Research in Galaxies and Cosmology, 
   University of Science and Technology of China, Chinese Academy of 
   Sciences, China}

\date{Accepted 2009 September 2.
      Received 2009 September 2;
      in original form 2009 April 30}

\pagerange{\pageref{firstpage}--\pageref{lastpage}}
\pubyear{2009}

\maketitle

\label{firstpage}

\begin{abstract}
Many brightest cluster galaxies (BCGs) at the centers of  X-ray selected clusters exhibit clear
evidence for recent star formation. However, studies of BCGs in optically-selected clusters  show that
star formation is not enhanced when compared to control samples of non-BCGs of similar stellar mass.
Here we analyze a sample of 113 BCGs in low redshift ($z<0.1$), optically-selected clusters, a
matched control sample of non-BCGs, and a smaller sample of BCGs in X-ray selected clusters. We convolve the SDSS images of the BCGs to match the
resolution of the GALEX data and  
we measure UV-optical colours in their inner and outer regions.
We find that optically-selected BCGs exhibit smaller 
scatter in optical colours and redder inner $NUV-r$
colours than the control galaxies, indicating that 
they are a homogenous population with very little
ongoing star formation. 
The BCGs in the X-ray selected cluster sample span a similar range in optical
colours, but have bluer $NUV-r$ colours. Among X-ray selected BCGs, those located in clusters with
central cooling times of less than 1 Gyr are significantly bluer than those located in clusters where
the central gas cooling times are long. 
Our main conclusion is that the location
of a galaxy at the centre of its halo is not 
sufficient to determine whether or not it is currently
forming stars. One must also have information about the 
thermodynamic state of the gas in the core of
the halo.
\end{abstract}

\begin{keywords}
 BCG; cooling flow; star formation
\end{keywords}

\section{Introduction}
\label{sec:intro} As their name already suggests, brightest cluster galaxies (BCG) are special in that
they are among the most luminous and massive
galaxies in the Universe and are  located near the minimum of the gravitational potential
of clusters and groups.
Observations show that although most BCGs
are elliptical galaxies they are different from ordinary galaxies of that type: they have surface brightness
profiles better described by two-component models \citep{Gonz03}, and their Fundamental-Plane
projections have  different slopes compared to normal ellipticals. This indicates
the formation histories of BCGs are likely to be different \citep{Anja07,Bern07,Desr07,Liu08}.

$N$-body simulations of BCG formation in a $\Lambda$CDM cosmology show that these galaxies form their
stars early ( $z$$\sim$5), but assemble their  final masses very late ( $z$$\sim$0.5) (\citet{Lucia07}
and references therein). The dominant stellar popultions of BCGs are 
indeed observed to be  old. Studies of the ultraviolet luminosities \citep{Hicks05,Pip08}, nebular line
emission \citep{Allen95,Craw99,Edw07}, optical absorption line spectra \citep{Card95}, and infrared
fluxes \citep{Egami06,ODea08} of BCGs show that a small amount of residual star formation is still
taking place in some of these objects. The existence of  UV -bright cores \citep{Bild07,Raff08,Pip08}
indicates that the star formation is centrally concentrated. In some cases, the star formation
and the radio jets coincide, possibly implying that the starbursts
are triggered by the jets 
\citep{Dona07a,ODea04}.

Star formation in BCGs is often associated with the so-called ``cooling flow'' phenomenon. Star forming
BCGs are often found nearer to the X-ray  centers of  clusters than quiescent BCGs
\citep{Bild07,Craw99,Edw07} and star formation is correlated with the cooling timescale of the gas.
Studies using X-ray data found that BCGs with star formation are clearly marked by an entropy threshold
of $\sim30$ KeV cm$^{-2}$ (equivalent to a cooling time scale of $\sim$1 Gyr and indicative of a strong
cooling flow \citep{Raff08,Cava08,Voit08}). In general, the mass of young stars in the BCG is only a
few percent of the gas mass that is predicted to be  condensing from the intracluster medium and
accreting onto the BCG. This may reflect the ability of clusters to distribute energy released by an
AGN to the surrounding gas \citep{Voit08}.

Around $70\%$ of clusters selected from flux-limited X-ray surveys are classified as cooling flow
clusters, and this high fraction holds to at least $z\simeq 0.4$ \citep{Edge97,Peres98,Bauer05}. If
this fraction is universal for all clusters, and enhanced star formation is directly related to the
cooling flow phenomenon, then BCGs should  form stars more actively than
ordinary massive elliptical galaxies. However, studies of BCGs in optically selected groups and
clusters have found that BCGs are as old as their field counterparts  \citep{Craw99,Anja07,Edw07}.

The aim of our study is to clarify whether BCGs are indeed forming stars more actively than other
massive elliptical galaxies.  We will make use of a combination of UV and optical imaging data.
UV/optical colors are sensitive to small amounts of star formation on timescales of $\leq$100 Myr. The
colors can be measured directly from images and will therefore not be affected by aperture  effects. In
contrast, the   spectral properties of  BCGs selected from the SDSS survey  are measured through a 3
arcsecond diamteter fibre aperture, and hence only probe star formation in the central regions of the
galaxies. The imaging data also allows us to study the colour gradients of the BCGs. We describe our
sample selection and our measurements of 2-zone colors in Section 2. We  present the comparison of
2-zone colors and color gradients for different BCG samples in Section 3. Finally, our results are
discussed in Section 4 and our conclusions are given in Section 5. Throughout this paper, we
assume a cosmology with $H_0$=70 km s$^{-1}$ Mpc$^{-1}$, $\Omega_m$=0.3,
and  $\Omega_\Lambda$=0.7
(\citep{Tegmark04}).
\section{Data and Sample}
\label{sec:sample}
\subsection{SDSS and GALEX }
\label{subsec:smp1}
 The Sloan Digital Sky Survey (SDSS) \citep{York00} has observed a quarter of the
extragalactic sky and provides images in five photometric bands (u',g',r',i',z'), as well as  spectra
of galaxies selected from the imaging. The spectra are taken with $3''$ diameter fibers and cover a
wavelength range from   3800 to 9100 ${\AA}$. The SDSS images have a pixel scale of $0.''396$ and a
mean PSF (point spread function) of $1.''4$ (FWHM).

The Galaxy Evolution Explorer (GALEX, \citet{Martin05}) is an orbiting space telescope providing
imaging in two bands: the far-ultraviolet (FUV) centered at 1528${\AA}$ and the near-ultraviolet (NUV)
centered at 2271${\AA}$. The images have $1.''5$ pixels and the average resolution is $4.''3$ FWHM for
FUV and $5.''3$ for NUV. GALEX  is performing a number of surveys distinguished mainly by coverage and
depth. In this work we use the data from the All-sky Imaging (AIS) with a typical depth of 20.5 in AB
magnitude and the  Medium Imaging survey (MIS) with a typical depth of 23.5 AB magnitude. We note that
a few NUV image tiles are not covered by FUV observations, but  all FUV images have accompanying NUV
images. Because the shape of the PSF near the border of GALEX images is distorted, we only use those
BCGs within 1200 pixels of the corresponding image center.

\subsection{``C4 Cluster Catalog'' and \citet{Anja07} catalog}
\label{subsec:smp2} \citet{Miller05} developed the ``C4'' algorithm to identify galaxy clusters from
the SDSS spectroscopic sample. They based their method on the fact that the cores of galaxy clusters
and groups are dominated by red  galaxies and used the SDSS colors($u-g$,$g-r$,$r-i$,$i-z$) for their
selection. The ``C4'' catalog is 90$\%$ complete for  galaxy clusters
with masses greater than
2$\times$10$^{14}$/h M$_{\bigodot}$ ( or velocity dispersion
$\sigma$ greater than 500 km/s). The completeness 
declines for lower mass clusters, reaching a value of $\sim$55$\%$ at a mass of
10$^{14}$/h M$_{\bigodot}$ ($\log \sigma_v$=2.6) and $\sim$30$\%$ at 2$\times$10$^{13}$/h
M$_{\bigodot}$ ($\log \sigma_v$=2.4) ( Here, we use Equation 2 from \citet{Biviano06} is used to convert cluster mass
to velocity dispersion ).

Von der Linden et al. (2007) (vdL07) based their study on the ``C4 Cluster Catalog'' of the Third Data
Release (DR3) of SDSS (748 clusters in total) and carefully identified the BCGs of every cluster. They
searched for BCG candidates within the virial radius and identified the BCG as the brightest galaxy
closest to the centre of the cluster potential . Their resulting sample consisted of 625 BCGs in groups
and clusters at $z\leq$0.1. They also recalculated the velocity dispersions and redshifts of each
cluster.

Our study is based on the vdL07 BCG sample and we refer to BCGs selected from this sample as ``optical
BCGs''. 402 of the vdL07 BCGs have SDSS spectra. We match the vdL07 sample with the  fourth General
Release of GALEX (GR4) and find   113 galaxies with MIS coverage in the NUV band (Sample 1, S1). We
prefer to use the MIS images rather than the shallower AIS images because the UV emission in most of
the BCGs is weak. 

As we will discuss, the majority of clusters in the
vdL07 sample are low in mass and have smaller  
velocity dispersions and X-ray luminosities
than typical clusters in X-ray samples (Figure~\ref{fig:sigv}). We thus created 
an additional sample of 60
massive optically-selected clusters  
with $\log \sigma_v$  greater than 2.8 and GALEX AIS images. 
We will use this sample when we compare optically-selected clusters with
X-ray selected clusters.

\subsection{Control sample}
\label{subsec:smp3} We constructed a control sample 
of field galaxies matched to the  S1 BCGs,  following the
steps outlined in vdL07. We  first sort the BCGs in mass  
and then we search the SDSS DR4 and
GALEX GR3 cross-matched catalogues  for non-BCGs that differ in 
redshift, stellar mass and $g-r$ color by less than 
0.02, 0.1 dex and 0.15 mag respectively. Each control galaxy enters the
sample only once.  

\begin{figure*}
\bc \hspace{0cm}
\includegraphics[width=.9\hsize]{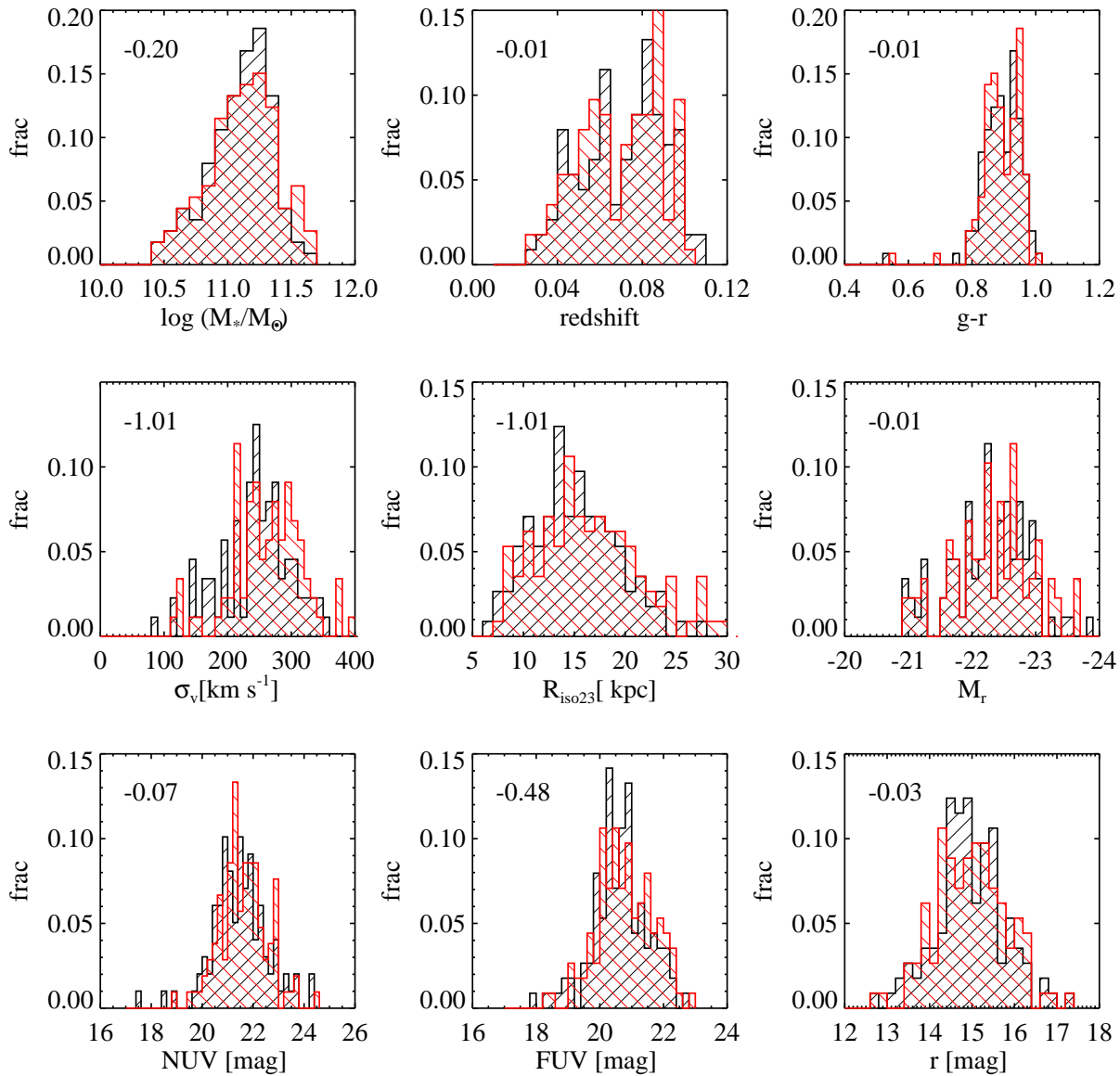}
 \caption {A comparison of the S1 BCGs with the control sample. We plot histograms of stellar mass, redshift, $g-r$ colour,
 galaxy velocity dispersion, R$_{iso23}$ radius, absolute r-band (AB) magnitude, and  NUV and FUV apparent magnitudes. BCGs are in red
and control galaxies are in black. In the top left-hand corner of each panel we list the logarithm of
the Kolmogorov-Smirnov probability that the two distributions are drawn from an identical parent
population (a 99 percent probability that the distributions are different will have a value of -2).}
\label{fig:control} \ec
\end{figure*}

Figure~\ref{fig:control} presents a  comparison of the UV and optical
properties of S1 BCGs and the control sample. 
Note that a significant fraction of the highest mass BCGs
lack comparable control galaxies; this is because almost all the very most massive galaxies in the
local Universe are themselves BCGs. We note that because the C4 cluster
sample is less complete at low masses, some of the control galaxies 
for these systems may themselves be BCGs. 

\begin{figure*}
\bc \hspace{-0.8cm} \resizebox{.9\hsize}{!}{
\includegraphics[width=.9\hsize]{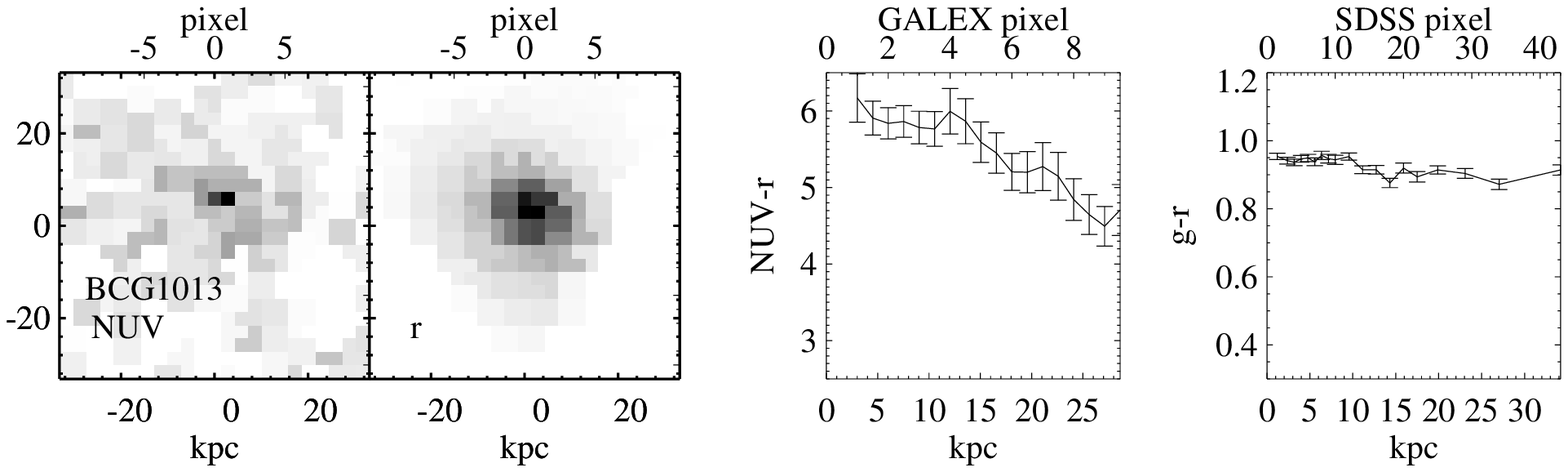}}\\[2pt]
\hspace{-0.8cm}\resizebox{.9\hsize}{!}{
\includegraphics[width=.9\hsize]{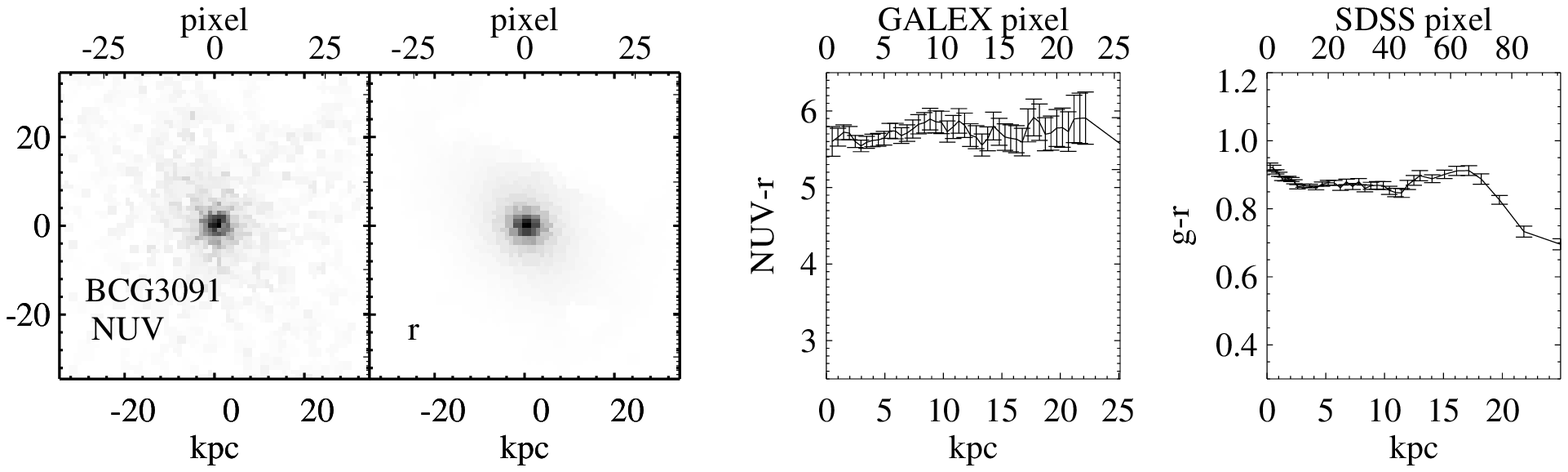}}\\[2pt]
\hspace{-0.8cm}\resizebox{.9\hsize}{!}{
\includegraphics[width=.9\hsize]{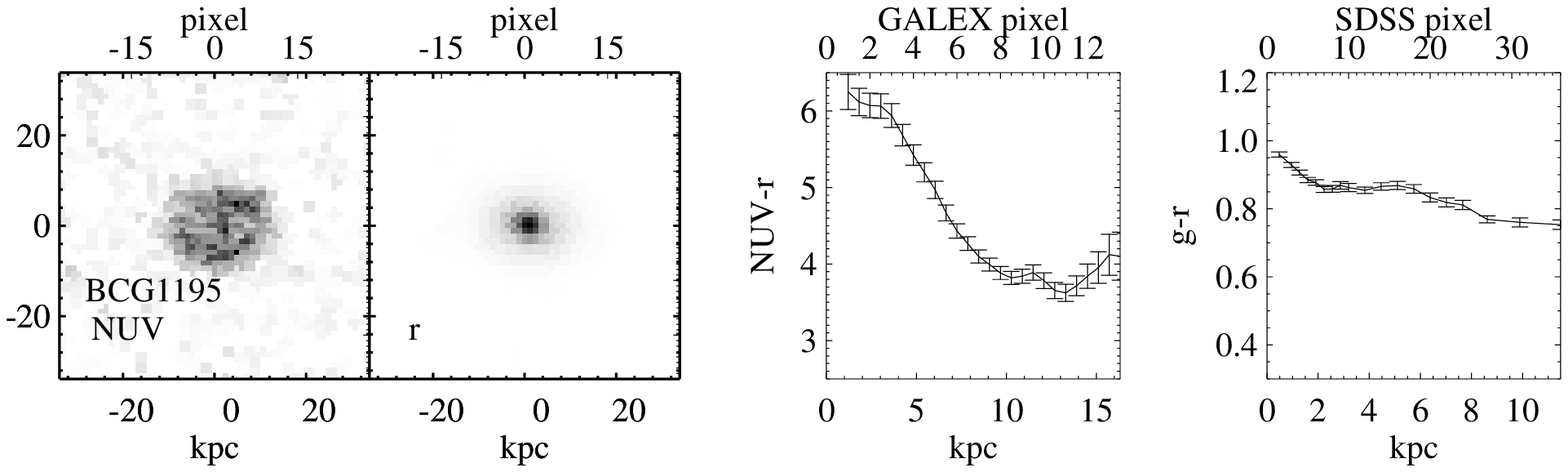}}\\[2pt]
\hspace{-0.8cm}\resizebox{.9\hsize}{!}{
\includegraphics[width=.9\hsize]{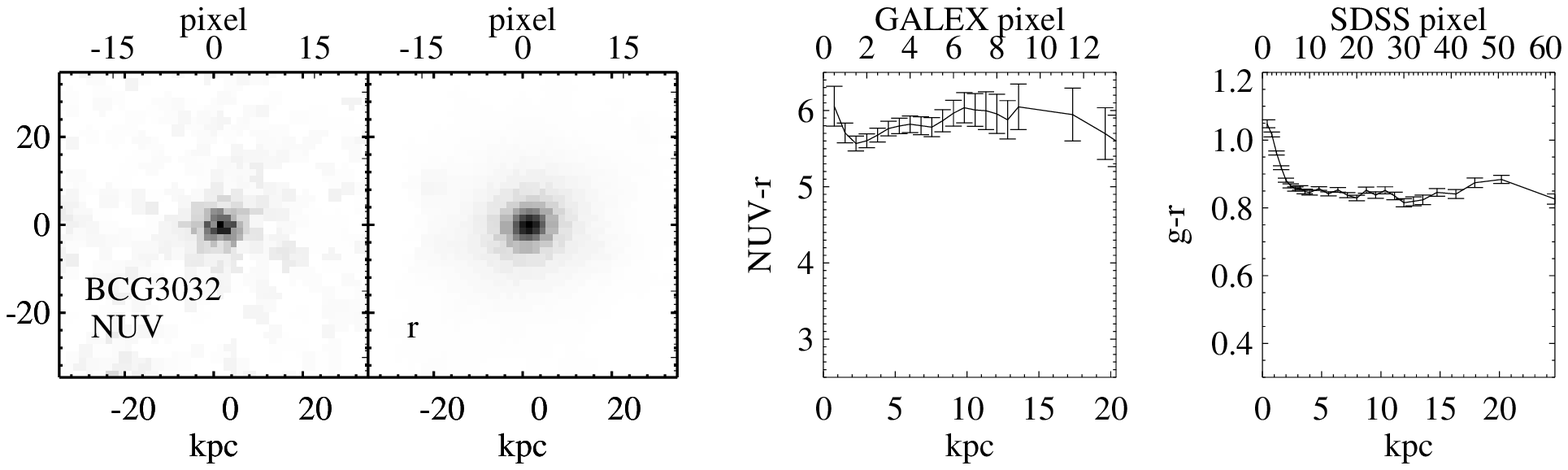}}
\caption{Some examples of typical  BCGs in our dataset. The first row shows a typical BCG in our sample
with low S$/$N in the UV, the second row  a UV bright BCG with red colour and flat $NUV-r$ gradient,
the third row a UV-bright BCG with a steep colour gradient, and the fourth row a BCG with a blue
$NUV-r$ core.} \label{fig:bcgsample} \ec
\end{figure*}

\subsection{X-ray BCGs from \citet{Raff08}}
\label{subsec:smp4} X-ray data provides information about the thermodynamic state of the gas in
clusters: by deprojecting the X-ray spectra extracted in elliptical annuli from X-ray images, one can
derive  gas temperatures and densities and thereby estimate gas  cooling time-scales and entropy.
Traditionally,  those clusters with ``central'' (determined by the limiting resolution of images)
cooling time-scales shorter than the age of the universe ($\sim$10 Gyr) have been classified  as
cooling flow clusters \citep{Fabian94}.  Recently, however, several studies \citep[][C08, R08
respectively afterwards]{Cava08,Cava08b,Raff08} have attempted to measure entropy and cooling times at
smaller cluster-centric radii using higher resolution data from the Chandra satellite. These studies
established a more stringent cooling time threshold of $\sim$1 Gyr (or 0.8 Gyr at a radius of 12 kpc
from \citet{Raff08}) below which BCGs were observed to have significantly enhanced star formation. In
our study, we adopt the latter definition, and refer to clusters (and their associated BCGs) with
central cooling times of less than 1 Gyr as cool-core clusters/BCGs, and those with central cooling
times greater than 1 Gyr as non-cool-core clusters/BCGs, respectively.

To assess whether X-ray and optically selected  BCGs differ, we make use of the sample of 46 BCGs
in \citet{Raff08} with X-ray data from Chandra. 
We search for optical images in
SDSS DR7 and find 21 matches (11 of these have spectra ): 
12 out of the 21 are in cool-core clusters
and 9 are in non-cool-core clusters. 
In order to compare colors$\backslash$stellar populations to the optically
selected BCGs, we need to further restrict the sample 
to lie at $z$$\leq$0.1 and have GALEX coverage, resulting in a total of only
9 clusters (3 non-cool-core and 6 cool-core) (see Table~\ref{T:X-ray BCG}).

\section{Photometry}
\label{sec:photo}
To ensure that our photometric measurements are consistent across the different bands, we  transform all our images
to the same geometry and effective resolution so that our photometric measurements trace the same
part of each galaxy at different wavelengths.
\subsection{Registering the images}
\label{subsec:pht1}
 We first run SExtractor$^1$\footnotetext[1]{http://sextractor.sourceforge.net/} on each image to get
 the celestial and
pixel coordinates for sources in the images \citep{Bertin96}. We then select the most compact sources
in the NUV images and match them to the celestial coordinates of sources in the SDSS and FUV images. We
run the GEOMAP/GEOTRAN$^2$\footnotetext[2]{http://iraf.noao.edu/} tasks in IRAF, with the pixel
coordinate pairs as input to register every image to the frame geometry of the corresponding NUV image.
\subsection{Convolving the images to a common PSF}
\label{subsec:pht2}
 We first need to obtain the PSF for each image. For SDSS images, we use the ``$read\_PSF$'' software$^3$
\footnotetext[3]{http://www.sdss.org/dr6/products/images/read$\_$psf.html} and read out the PSF models
for the five bands directly from the corresponding psField file. We transform these SDSS PSF models to
a GALEX pixel scale of $1.''5$. In order to get the PSF model for the NUV and FUV images, we first
select stars within a radius of 1200 pixels around the center of each image. After masking nearby
sources, we fit a two-dimensional gaussian function to the stars. We cut out small stamps around each
star and shift them so that every star is centered exactly on the center of the stamp. We add these
stamps together and adopt the combined stack as the best PSF model for the frame. We check the PSF
models by subtracting the stack from each of the input stars. Then we run the IMMATCH/PSFMATCH task in
IRAF to obtain kernel functions that we use to convolve the higher resolution SDSS images to the
resolution of the corresponding NUV images. Figure~\ref{fig:bcgsample} shows examples of the registered
and convolved images of 4 different  S1 BCGs with different UV/optical colours. The resulting images
are adaptively binned in two dimensions using the algorithm of \citet{Cap03} so that all pixels have a
signa-to-noise ratio ($S/N$) above a certain fixed value. The resulting images are then able to show
the extended, low surface brightness regions of the sources. The color profiles are obtained by
averaging the colors in a set of elliptical rings with increasing radius around the BCG center. The
rings are chosen to that all all points along the profile have errors below a certain fixed value. The
first row in Figure~\ref{fig:bcgsample} shows a typical BCG in our sample with low S$/$N in the UV, the
second row shows a  UV-bright BCG with red colors and a flat $NUV-r$ profile, the third row shows a
UV-bright BCG with a steep  color profile, and the fourth line a BCG with a blue $NUV-r$ core.

\begin{figure}
\bc \hspace{-0.8cm} \resizebox{6cm}{!}{\includegraphics{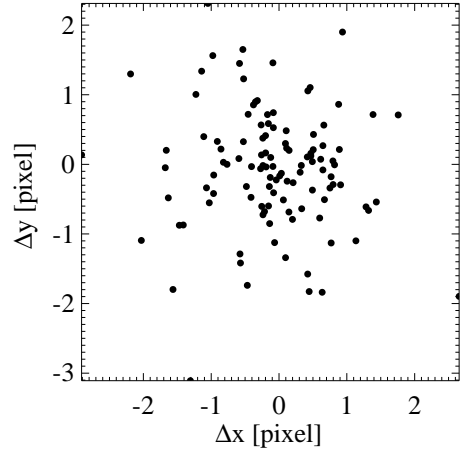}} \caption{Differences between UV
and R-band  centroid positions (in GALEX pixel units) of the BCGs in our sample. }
 \label{fig:t1} \ec
\end{figure}

In Figure~\ref{fig:t1}, we plot the positional differences between the centroid of the  BCG  measured
from the convolved SDSS and GALEX images. As can be seen, the centroids generally agree to better than
$1''$, but in a few cases the offsets can be as large as 3 arcseconds. We have checked these images by
eye and  we find the largest offsets between the centroids can be caused by (1) SExtractor centroiding
on regions of star formation, which are more prominent in UV light, (2) failure of SExtractor to
deblend the BCG from nearby UV bright sources, or (3) low $S/N$ of the UV image. So it is important to
use SDSS image decided position parameters for both GALEX UV and SDSS photometry to ensure consistency.


\subsection{Two-zone photometric measurements}
\label{subsec:pht3} In order to assess whether the BCGs in our sample have colour gradients, we measure
colours both inside and outside a fixed aperture for each object in our sample. Because the SDSS $r$
band images are of good quality and high $S/N$ ratio, we use them as the basis for our two-zone
photometry. We first mask all sources except the BCG. We use SExtractor to determine the ellipticity
and position angle from the $r$ band image \citep{Bertin96}. These determine the  shape and position
angle of the apertures that we use. Following vdL07, we define the total  magnitude of the BCG as the
light contained within the radius  where the $r$-band surface brightness equals $(23+10 \log(1+z))$ mag
arcsec$^{-2}$ (which we denote as $r_{iso23}$). All our photometry measurements are processed inside
this aperture, which encloses most of the light. The reason why we measure the BCG magnitude within
$r_{iso23}$ and do not attempt to measure a total magnitude, is because the surface brightness profiles
of BCGs are complex and do not follow a simple de Vaucouleurs profile \citep{Gonz05}. We define
R$_{50}$ and R$_{90}$ as the radii enclosing 50\% and 90\% of the total flux of the galaxy evaluated
within $r_{iso23}$. We measure the total UV and $r$-band  flux inside R$_{50}$ (in), and between
R$_{50}$ and R$_{90}$ (out). We compute colours within these two regions and and define the color
difference, $\Delta(NUV-r)=(NUV-r)_{\mathrm in} - (NUV-r)_{\mathrm outs}$. We apply this definition to our
UV/optical colour measurements, because it has higher $S/N$ than more traditional measures of colour
gradient. We caution that in Sect.~\ref{subsec:pht5} we will use a different measure of colour
gradient, because we are comparing directly with existing studies published in the literature.

\begin{figure*}
\hspace{-0.8cm} \resizebox{12cm}{!}{
\includegraphics{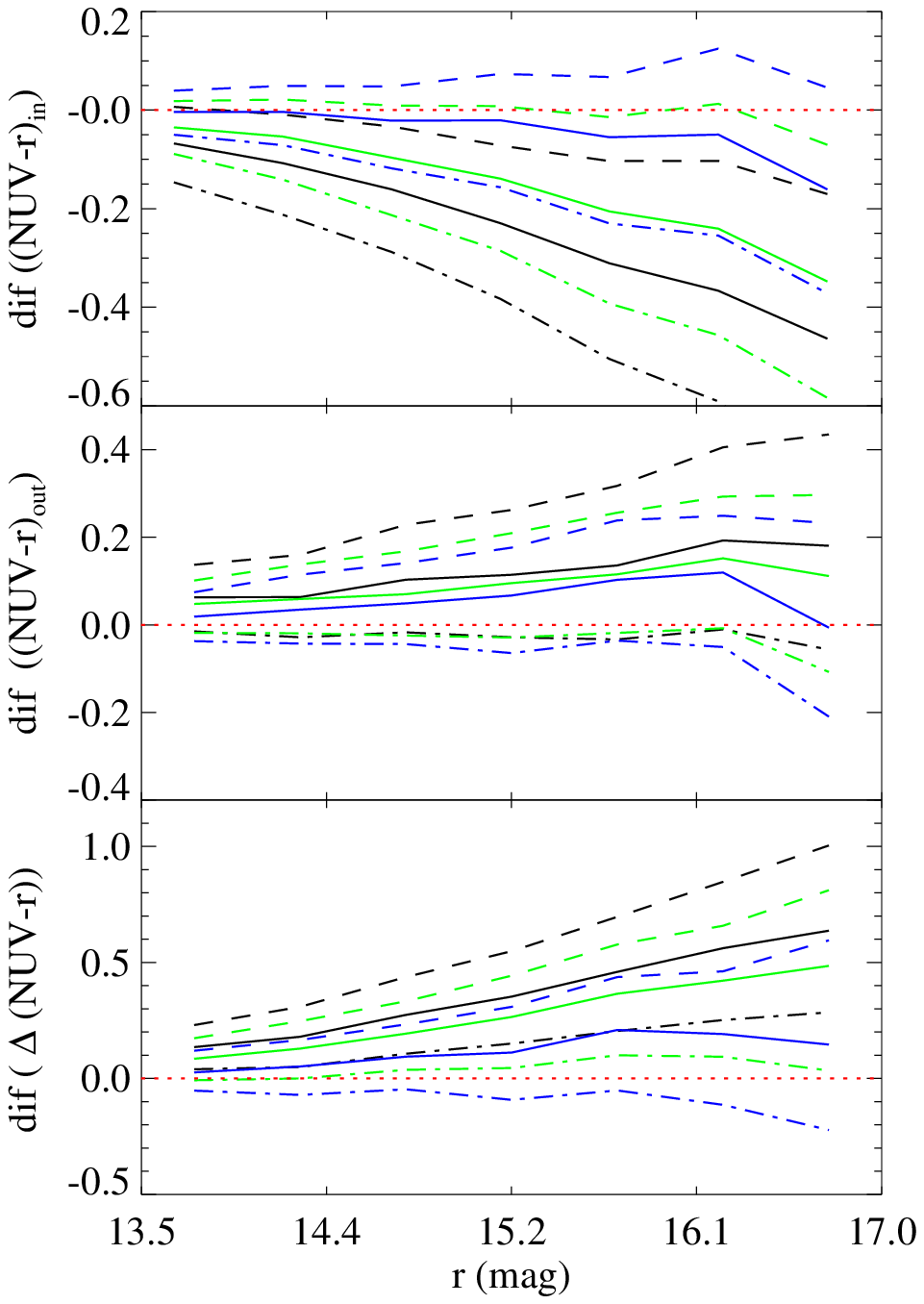}
\hspace{0.1cm}
 \includegraphics{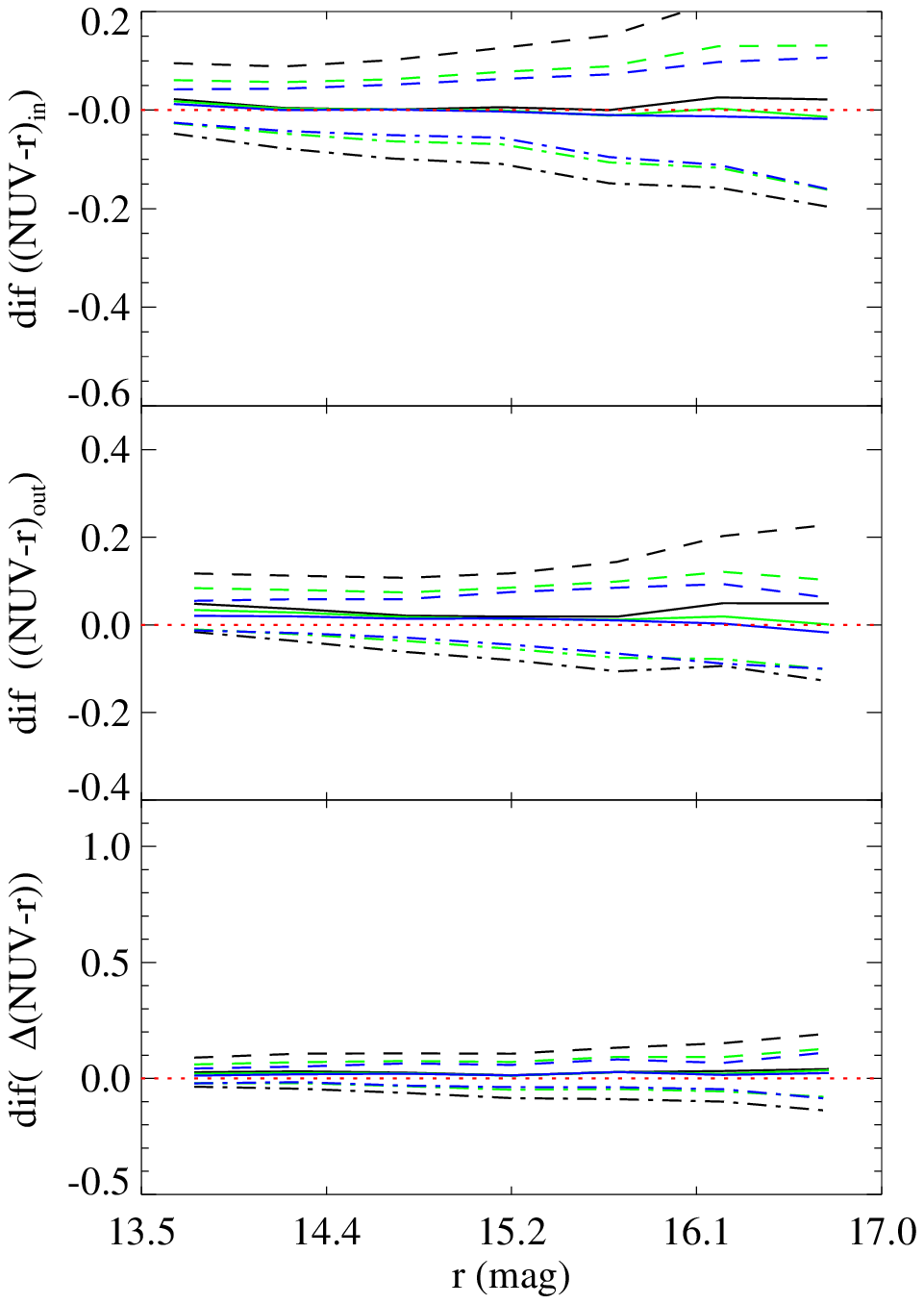}}
\caption{ The results of tests using simulated images showing how the convolution procedure and sky
noise will affect our colour measurements.  The left column shows differences between the intrinsic
(input) colours  and the colours measured from the  simulated images, which have been convolved to the
resolution of the GALEX data and which have Poisson noise added. The right column shows the effect of
just adding the Poisson noise. The data are divided into three different color bins
([5,5.5],[5.5,6],[6,6.5]) ( blue, green and black lines in the plot). Solid lines show the median
values at a given $r$-band magnitude, while dashed lines show the $25\%$ and $75\%$ percentiles of the
distribution at the same magnitude  (see text for details).} \label{fig:t4}
\end{figure*}

\begin{figure*}
\hspace{-0cm} \resizebox{12cm}{!}{
\includegraphics{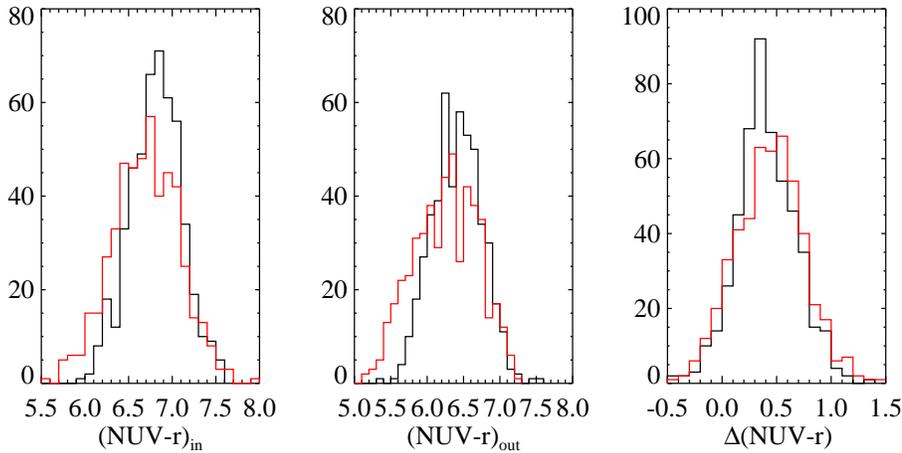}}
\caption{ An analysis of the influence of structural parameter differences between BCGs and non-BCGs
on our  2-zone color measurements (see text for details). Results are shown for simulated BCGs (red)
and simulated non-BCGs (black). The simulated BCGs and non-BCGs have the same distribution of colours.
The colour profiles of both kinds of objects are assumed to be flat. The differences between the red
and black histograms arise purely because of the structural differences between the two classes of
object.} \label{fig:t5}
\end{figure*}

\subsection{Accuracy of our 2-zone color measurements}
\label{subsec:pht4} Recall that our two-zone photometry is carried out on
optical images that have been convolved to the resolution of the GALEX
data. It is thus important to understand  to what extent our
colour difference measurements reflect the intrinsic
values for the real galaxy.

To assess this, we simulated 10,000 galaxies in the $r$ and NUV bands and included a similar level of
background noise as in our real data (e.g. the MIS background for the simulated NUV images). The models
assume Sersic profiles with $n$=4 for the $r$ band  and $n$=2 for the UV band
(i.e. we assume that the optical light comes from a bulge-like component
and the UV from a star-forming disk). The range in
$r$ magnitude, NUV magnitude and $r$-band effective radius for the simulated galaxies are constrained
to be the same as in the vdL07 sample. By varying the effective radius of the NUV light, we obtain a
set of simulated galaxies with different colour gradients. We then apply our algorithm to these images
and compare our colour measurements to the intrinsic "input" values in Figure~\ref{fig:t4}. The inner
color is defined to be the color inside R$_{50}$ and the outer color is defined as the colour  between
R$_{50}$ and R$_{90}$.

We begin by exploring the effect of noise of our colour measurements.  The right column in
Figure~\ref{fig:t4}  shows the differences between the measured colours for  model images with and
without Poisson noise. The median value of the differences is zero across most of the  $r$-magnitude
range spanned by our BCG sample. Systematic effects only become important at  $r$ magnitudes fainter
than 16.1 (most of our BCGs are brighter than this). We then investigate what effect the convolution to
GALEX resolution has on our measurements. The left column of the plot shows the measurement differences
between model images with perfect resolution and the GALEX NUV resolution. The top panel shows that the
inner $NUV-r$ colour will be underestimated and the situation is worse for redder objects. The middle
panel shows that the  outer $NUV-r$ colour will be over-estimated and the situation is again worse for
redder objects. However, the   outer color is better constrained in general  than the inner colour. At
an $r$-band magnitude of $\sim$16.0, the inner $NUV-r$ of the reddest objects will be underestimated by
$\sim$0.3 mag, while that of the bluest ones would be underestimated by  $\sim 0.1$ mag. The outer
$NUV-r$ colour would be over-estimated by $\sim$0.15 mag on average. The bottom panel shows that the
resulting color difference $\Delta(NUV-r)$ will be over-estimated by  $\sim$0.5 mag at $r\sim$16.0.
These results are easily understood from the fact that the optical light (with Sersic model $n$=4) is
more concentrated than the UV light (with Sersic model $n$=2), and will thus be more severely scattered
to the outer region of the galaxy. As we will show later on, our results are not substantially affected
by these systematics.

Von der Linden et al.(2007) found that BCGs have larger effective radii and lower surface brightnesses
than non-BCGs. We simulated two sets of galaxies with structural parameters similar to the optical BCGs
and the control sample to check the extent to which differences in structural parameters
 affect our measurements of 2-zone colors. We fix the $r$ band
magnitude to be 15 ($\sim$ median value of the optical sample), $NUV-r$ to be 6.0 (a typical value at
the red end of our sample) and $\Delta(NUV-r)$ to be 0. Only the mean surface brightness inside
$R_{50}$, $\mu_{50}$, is allowed to vary ($R_{50}$ will vary accordingly since the total $r$ magnitude
is fixed). The median values of $\mu_{50}$ are fixed to be 18.82 and 18.65 mag for BCGs and non-BCGs
respectively (corresponding to the median values of the two samples as measured by vdL07). $\mu_{50}$
is  allowed to vary randomly around the median value with a scatter of 0.7 mag arcsec$^{-2}$ (again
according to von der Linden's distribution). We compare  the distributions of the values of
$(NUV-r)_{\mathrm in}$, $(NUV-r)_{\mathrm out}$, and $\Delta(NUV-r)$ (defined as in Sect~\ref{subsec:pht3}) for BCGs
(red) and non-BCGs (black) in Figure~\ref{fig:t5}. It can be seen that structural differences cause
both the inner and outer colours of BCGs to be shift slightly bluewards relative to non-BCGs.  However,
the color {\em differences} for the two samples are similar. We have also  simulated  cases where the
input galaxies have an  intrinsic colour gradient, and the results are similar.

\subsection{$u-r$ gradients}\label{subsec:pht5}
 The SDSS optical photometry is of higher resolution than the GALEX photometry, so
$u-r$ colour gradients offer an alternative way to trace star formation in the central regions of early
type galaxies.  As was found by R08 and will be seen in Sect~\ref{subsec:rst3}, X-ray selected
cool-core BCGs have steeper $u-r$ colour gradients than non-cool-core BCGs. We will use $u-r$ gradients
to link our study of optical selected BCGs to previous studies of X-ray selected BCGs (e.g. R08).

We use the SDSS $r$ and $u$ band images to measure the $u-r$ colour profiles. These images have lower
resolution and poorer seeing than the images used by R08, but we produce high quality color profiles by
registering and convolving the $u$ band images to the $r$ band images.  We then measure the color
gradients ($G(u-r)$) following the procedures oulines in  in R08.

\begin{figure}
\hspace{0.1cm} \resizebox{8cm}{!}{
\includegraphics{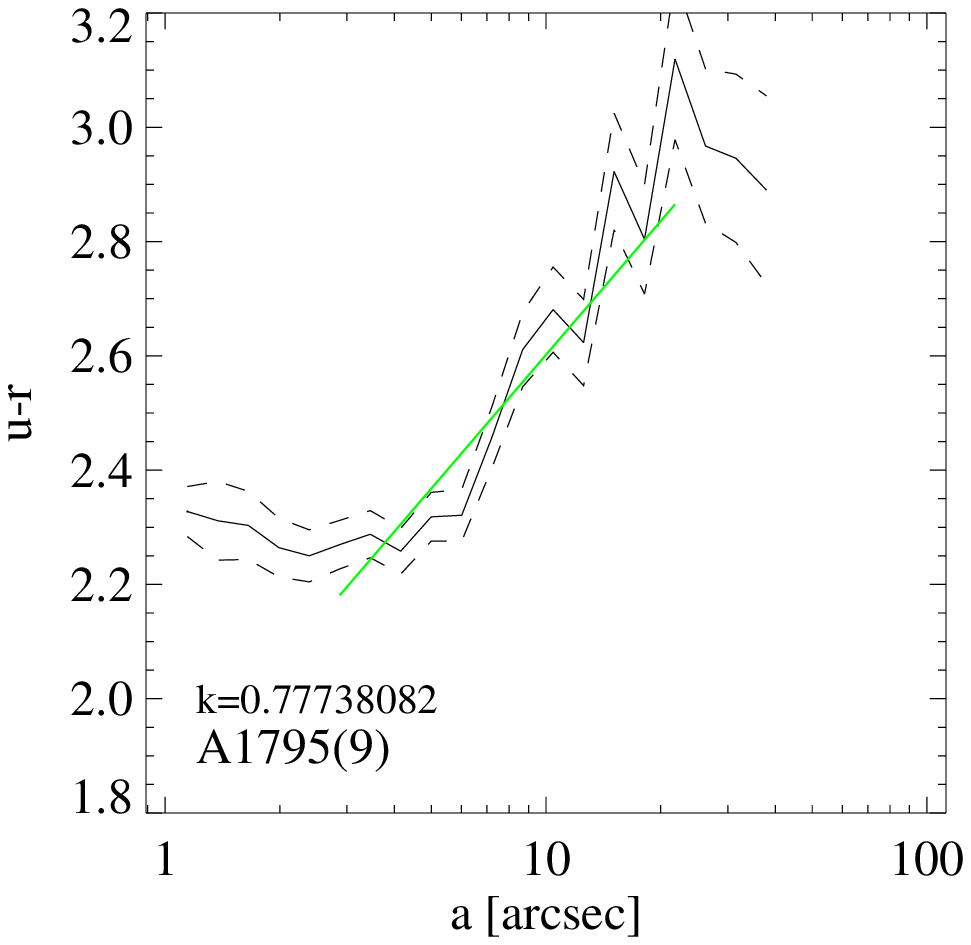}
\hspace{1cm}
\includegraphics{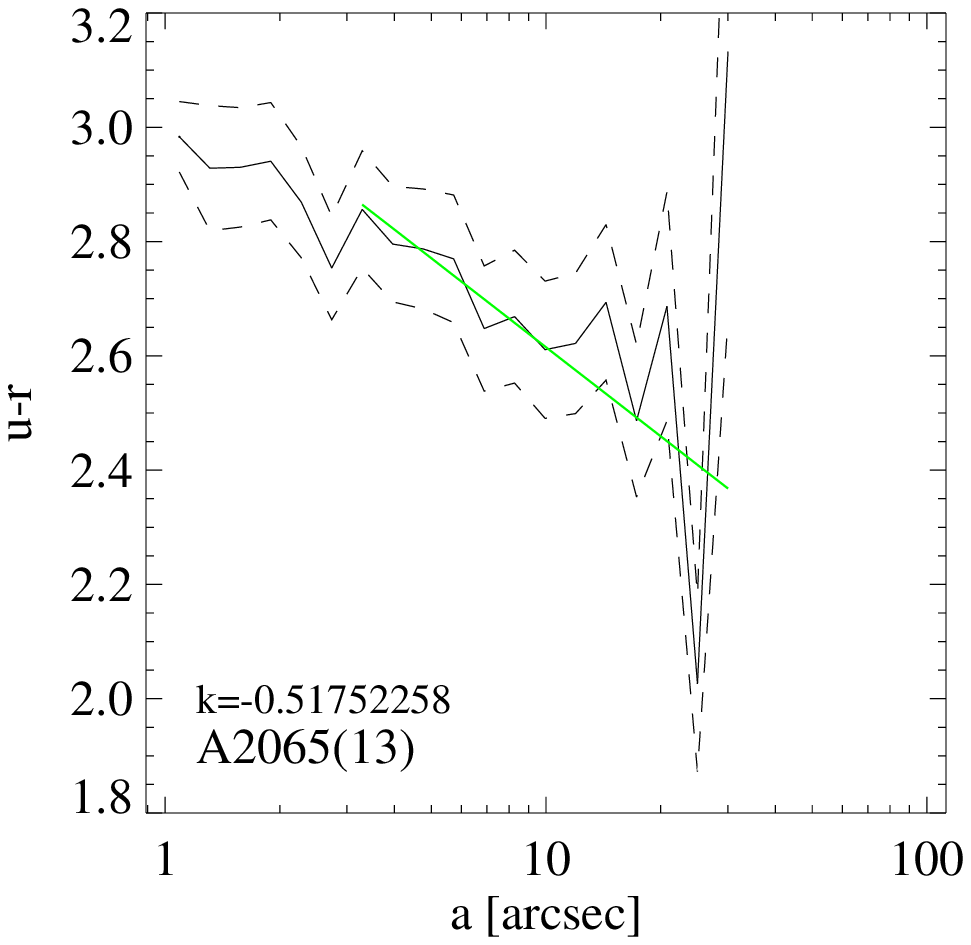}}
\caption{$u-r$ differential color profiles of two BCGs from the R08 X-ray sample (A1795 and A2065). The
dashed lines show the 1 $\sigma$ range of the profile. The best-fit gradients are overplotted as green
straight lines in the fitting ranges as defined in Sect.~\ref{subsec:pht5}. The values of the best-fit
gradients are shown in the left-down corner of each profile.} \label{fig:t6}
\end{figure}

The procedure we employ is the following. We measure colours in a series of  elliptical annuli, with
the major axis increasing in   $\sim$ 1 pixel intervals. If the $u-r$ profile has a positive color
gradients near the center that extends for more than 4 sampling intervals ($\sim$ 2.2 arcsec) from a
point near the center of the galaxy at twice the FWHM of the PSF, the galaxy is classified as having a
blue core.  The blue core region is defined to end  where the colour gradients become negative.
Otherwise, the galaxy is classified as having a red core. We  fit the colour profile in the radius
range from twice the FWHM of the PSF to where the excess blue emission ends (for objects with blue
cores) or where the total errors reach 0.5 mag (for objects with red cores), and define the colour
gradient $G(u-r)$ as the slope of the fit.
 Figure~\ref{fig:t6} shows the $u-r$ profiles and 
the range of our adopted  linear
fits for two BCGs from the R08 X-ray sample.

\begin{figure*}
\bc \hspace{-0.8cm} \resizebox{14cm}{!}{\includegraphics{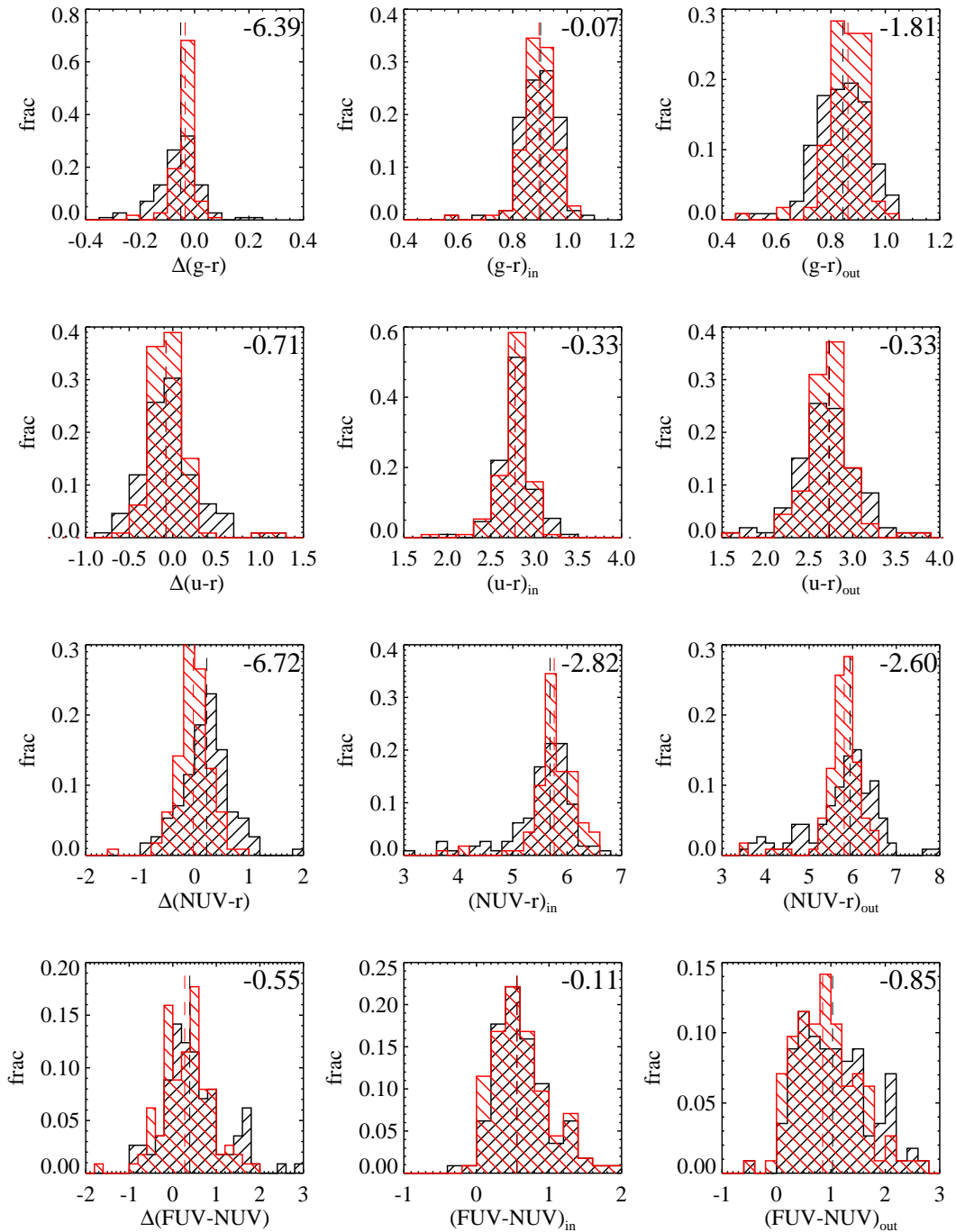}} \caption{Distribution of
2-zone colours and  colour differences for optical BCGs (red) and the control sample of massive field
galaxies (black). The dashed lines mark the median values for the two distributions.} \label{fig:bcg_c}
\ec
\end{figure*}

\section{results}
\label{sec:results}
\subsection{BCGs and control galaxies}
\label{subsec:rst1} In Figure~\ref{fig:bcg_c}, we compare the normalized distribution of colours and
colour differences for S1 BCGs (red) (see definition in Sect~\ref{subsec:smp2}) and control galaxies
(black). Colours are measured in two zones: inside $R_{50}$ (denoted by $colour_{\mathrm in}$) and between
$R_{50}$ and $R_{90}$ (denoted by $colour_{\mathrm out}$). Colour differences here follow the definition in
Sect 2. The results of a Kolmogorov-Smirnov test  (specifically,  the logarithm of the K-S probability
that the two distributions are drawn from an identical parent population) are given in each panel.
Smaller values mean larger differences between the two distributions (a 99 percent probability that the
distributions are different will have a value of -2). Dashed lines in each panel indicate  the median
values of the distributions.

The top two panels show SDSS colours measured for the BCGs and control galaxies.  There is no
systematic difference in median colour between the two samples. However, non-BCGs clearly  have larger
scatter in their colours than BCGs, particularly in their outer regions. This translates into a broader
range in colour differences, which is consistent with the conclusions of vdL07, who found that BCGs
have more homogeneous stellar populations than non-BCGs.

There are clear systematic shifts in the $NUV-r$ colours of BCGs and non-BCGs in the sense that BCGs
have bluer outer $NUV-r$ colours and redder inner $NUV-r$ colours. The first question one might ask is
whether these shifts are real, or whether they simply reflect that fact that BCGs have larger sizes
($R_{50}$) and lower optical surface brightness than non-BCGs. As discussed in Sect.~\ref{subsec:pht4},
this  may indeed partly explain why BCGs have bluer outer $NUV-r$ colours. Using the colour
distributions of our sample of BCGs and non-BCGs, we estimate that the net shift due to structural
differences between the two kinds of galaxies is about 0.1 mag, and thus cannot explain  the
differences between the two distributions (the reddest non-BCGs are $\sim$0.4 mag redder than the
BCGs). Also, we find that the inner $NUV-r$ colours of  BCGs are redder than those of non-BCGs, which
goes in the opposite direction to the effect predicted from the simulations In summary, our result that
BCGs have bluer outer $NUV-r$ colours than non-BCGs may be weaker than suggested in
Figure~\ref{fig:bcg_c}, but the result that  BCGs have  redder inner $NUV-r$ colours than non-BCGs is
robust. This would  imply that the inner stellar populations of  BCGs are older, more metal rich, or
more dusty (or some combination of these).

We find no detectable differences in the $FUV-NUV$ colours of BCGs and non-BCGs. This may be due to
lower quality of the FUV images. 

We note that the  BCGs and comparison galaxies have been matched in redshift, so
$K$-corrections are not an issue in our analysis.
Finally, as discussed in Sect~\ref{subsec:smp3}, the control galaxies correpsonding
to BCGs in less massive clusters may be "contaminated'' with BCGs.
We have checked whether there is any trend in the differences betweeen
the colours of BCGs and non-BCGs as a function of cluster mass, but
we did not find any effect.

\subsection{Optical BCGs versus X-ray BCGs}
\label{subsec:rst2} 
\begin{figure*}
\bc \hspace{-0.8cm} \resizebox{14cm}{!}{\includegraphics{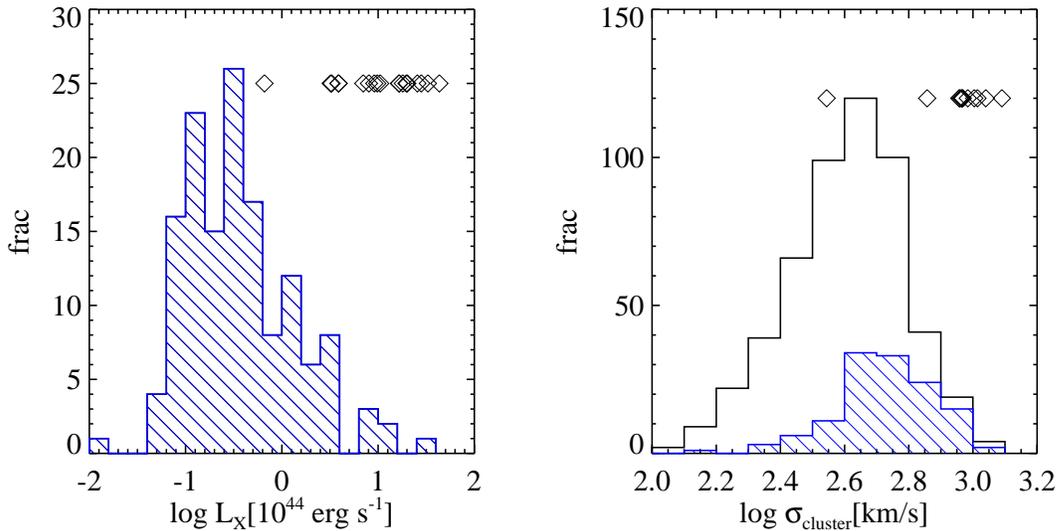}} \caption{$\sigma_{cluster}$
(right panel) and X-ray luminosity (left panel) of the entire optical BCG sample (black) and subset of
the optically-selected BCGs with data from  ROSAT (blue). The diamonds mark the values of the
X-ray selected R08 BCGs used in this work.} \label{fig:sigv} \ec
\end{figure*}

Figure~\ref{fig:sigv} shows the distributions of $\sigma_{cluster}$
and X-ray luminosity (when it is
available from ROSAT, \citet{Shen08}) for the vdL07 sample. In the following analysis, we limit the
optical BCG sample to those objects in clusters with $\log{\sigma_{cluster}}>2.8$ km s$^{-1}$. After
including galaxies with GALEX AIS photometry, the  sample consists of
 60 BCGs.
The X-ray selected BCGs from R08 (Table~\ref{T:X-ray BCG}) are still
significantly biased
towards higher  $\sigma_{cluster}$ and even more severely 
towards higher X-ray luminosities compared to the optically-selected
clusters. We will come
back to this point later in our  discussion.

\begin{figure*}
\bc \hspace{-0.8cm} \resizebox{16cm}{!}{\includegraphics{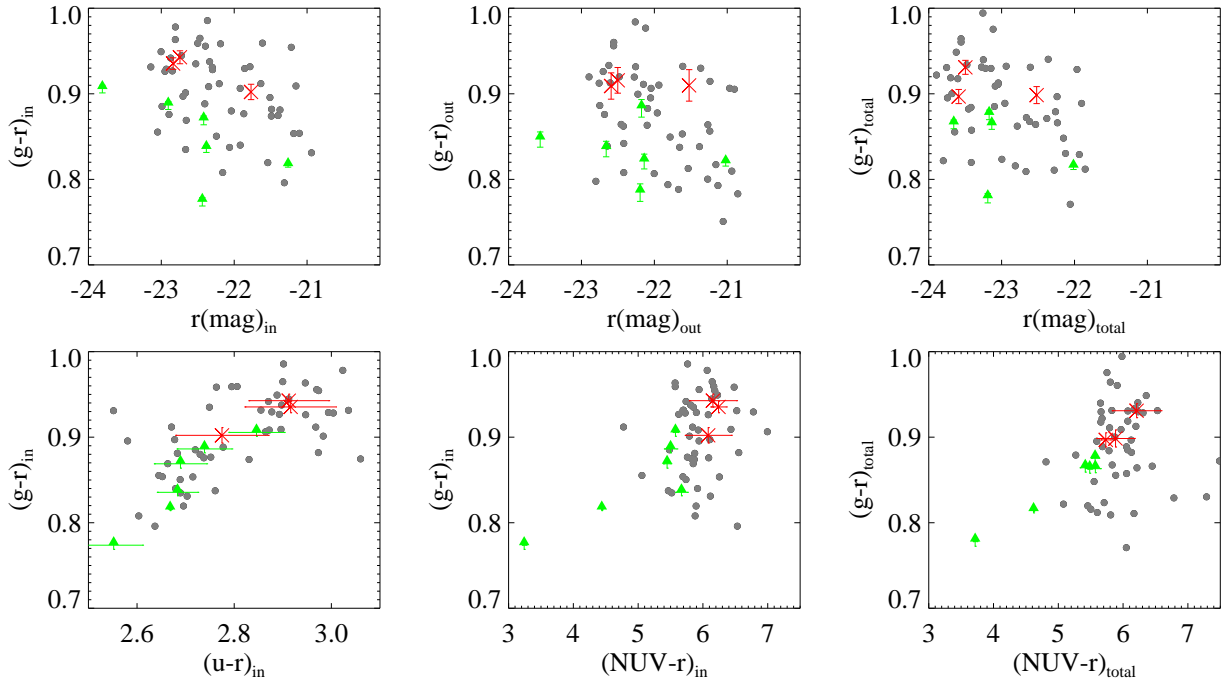}} \caption{Comparison between the
inner and outer colours of optical BCGs (black, filled circles; only those in clusters with
$\sigma_v>$2.8 are plotted), cool-core (green, filled triangle) and non-cool-core BCGs (red, crosses).
} \label{fig:opt_x} \ec
\end{figure*}

Figure~\ref{fig:opt_x} plots                           
the BCGs at z$\leq$0.1 in a series of colour/magnitude
and colour/colour diagrams. BCGs in optically-selected  clusters are shown as
black points and BCGs in X-ray selected clusters are shown 
as coloured symbols (cool-core BCGs in green and non-cool-core BCGs in red). 
Most X-ray selected BCGs have high $r$-band
luminosities compared to the optical BCGs. 
X-ray selected BCGs have similar $g-r$ and $u-r$ colours to the
optically selected BCGs, but blue inner NUV--r colours.
Among the X-ray selected BCGs, 
the  cool-core BCGs are always bluer than the
non-cool-core ones 
(cool-core and non-cool-core BCGs separate at a $NUV-r$ colour of $\sim$5.8). 
This demonstrates  that the UV
and optical colours of the X-ray selected BCGs are indeed correlated with the cooling time of the gas.
One of the objects with the bluest $NUV-r$ colour is the BCG in the well studied very strong cooling
flow cluster A1795 \citep{ODea04,Mittaz01}. 
The other blue outlier is M87, which is an AGN. The
synchrotron emission from its jet is quite prominent in the UV \citep{Madrid07}.

In Table 2, we list K-S test probabilities for colour differences 
between optically and X-ray selected BCGs, as well as optically selected
BCGs and cool-core BCGs. The results confirm our assertion that the most
significant differences are found for UV/optical colours and for
optically-selected and cool-core BCGs. Could the large differences in
UV/optical colours be an artifact of our convolution algorithm? 
As discussed in Sect~\ref{subsec:pht4}, the inner $NUV-r$ colours are more severely underestimated for
red objects than for blue objects. So if we were to correct our colour measurements to the value before
convolution, red galaxies  will become redder in $NUV-r$ while blue galaxies will not change much. In
other words, the $NUV-r$ colour differences between the cool-core and non-cool-core BCGs would be
enhanced even more.

\begin{table*}
\begin{center}
\renewcommand{\arraystretch}{1.3}
\begin{tabular}{l l c c c c c c c}
\hline \hline
 \multicolumn{1}{c}{Number} & \multicolumn{1}{c}{Name} & \multicolumn{1}{c}{Redshift} &
 \multicolumn{1}{c}{$\sigma_{cluster}$} &
\multicolumn{1}{c}{L$_X$} & \multicolumn{1}{c}{H$\alpha$} & \multicolumn{1}{c}{t$_{cool}$} &
\multicolumn{1}{c}{G(u-r)$_{R08}$}&
\multicolumn{1}{c}{G(u-r)} \\
\multicolumn{1}{c}{} & \multicolumn{1}{c}{} & \multicolumn{1}{c}{} &\multicolumn{1}{c}{(km s$^{-1}$)} &
\multicolumn{1}{c}{(10$^{44}$ ergs$^{-1}$)}& \multicolumn{1}{c}{(10$^{40}$ erg s$^{-1}$)} &
\multicolumn{1}{c}{(10$^8$ yr)} & \multicolumn{1}{c}{}&
\multicolumn{1}{c}{}\\
\hline
     1&       A85 & 0.055& 1097$^w$ & 16.5$^w$  &  0.502  & 6.4   &  $0.02\pm0.04$  &  $0.244\pm0.230$ \\
     2&      A383 & 0.187& 920$^w$ &  9.8$^s$  &    --  & 4.4   &  $0.15\pm0.28$  &  $-0.248\pm2.120$\\
     3&   Perseus & 0.018& 925$^w$ & 19.7$^w$  &    --  & 5.5   &     --          &  $0.568\pm0.010$ \\
     4&     A1361 & 0.117& --  &  3.22$^e$ &  13.5$^c$ & 8     &  $-0.15\pm0.06$ &  $-0.403\pm0.131$\\
     5&     A1413 & 0.143& 1231$^w$ &  16.6$^w$ &    0   & 15    &  $-0.29\pm0.15$ &  $-0.534\pm0.316$\\
     6&       M87 & 0.004& 350$^w$ &  0.66$^w$ &    --  & 8.6   &  $-0.27\pm0.08$ &  $-0.068\pm0.004$\\
     7&     A1650 & 0.084& 1033$^w$ &  7.01$^w$ &    --  & 15.5  &  $-0.14\pm0.11$ &  $-0.063\pm0.064$\\
     8&      Coma & 0.023& 1010$^w$ &  9.14$^w$ &  0.0298  & 71.8  &  $-0.14\pm0.02$ &  $0.018\pm0.016$ \\
     9&     A1795 & 0.063& 920$^w$ &  20.1$^w$ &  11.3$^c$ & 7.2   &  $0.5\pm0.1$    &  $0.777\pm0.072$ \\
     10&     A1991 &0.059&  720$^w$ &  3.29$^w$ &   1.1$^c$ & 4.1   &  $-0.35\pm0.06$ &  $-0.134\pm0.052$\\
11&  MS1455.0+2232 &0.258& 964$^H$ & 10.51$^{eg}$& 1.816  & 3.9   &  $-0.06\pm0.23$ &  $-0.557\pm1.271$\\
12& RXCJ1504.1-0248 &0.215&  --  & 28.07$^b$ & 475.595 & 2.9   &  $0.94\pm0.17$  &  $1.443\pm0.280$ \\
13&          A2065 &0.073& 908$^w$  &  3.88$^w$ &    --  & 15    &  $-0.38\pm0.08$ &  $-0.518\pm0.132$\\
14&  RXJ1532.8+3021 &0.354&  --  &  32.9$^h$ &   415$^c$ & 3.7   &  $1.04\pm0.4$   &  $0.209\pm0.532$ \\
     15&     A2244 &0.097& 930$^p$  &   6.6$^p$ &  0.333 & 17.4  &  $-0.2\pm0.12$  &  $0.278\pm0.161$ \\
     16&   NGC6338 &0.027&  --  &43.28$^{H1}$ &  0.909 & 9     &     --          &  $0.001\pm0.017$ \\
17&  RXJ1720.2+2637 &0.164&  --  & 25.58$^h$ &  12.7$^c$ & 5.3   &  $0.28\pm0.14$  &  $0.924\pm0.344$ \\
18& MACSJ1720.2+3536&0.391&  --  &    --  &    --  & 3.6   &  $0.84\pm0.85$  &        --       \\
19&          A2261 &0.224&  --  & 18.06$^e$ &   1.318  & 9.3   &  $-0.42\pm0.19$ &  $-0.578\pm1.194$\\
20&          A2409 &0.148&  --  &  8.04$^e$ &    0$^c$  & 23.7  &  $-0.34\pm0.26$ &  $-1.705\pm1.646$\\
21&          A2670 &0.076& 908$^w$& 3.88$^w$ &   0.1$^c$ & 23.4  &     --          &  $-0.694\pm0.085$\\
\hline \hline
\end{tabular}
\caption{Properties of the BCGs in the X-ray selected sample of (R08). Columns 2 and 3 list the
velocity dispersions and X-ray luminosities of  the host clusters, which have been obtained from a
variety of different sources as indicated by a
superscript to the values of $L_X$ as follows:
 $s$ refers to \citet{Smith01}, $e$ refers to \citet{Ebeling96}, $eg$
refers to \citet{Egami06}, $h$ refers to \citet{Hashimoto07}, $H$ refers to \citet{Hoekstra07}, $H1$ refers
to \citet{Helsdon01} , $p$ refers to \citet{Peres98}, $w$ refers to \citet{DAWhite97}. Column 4 lists the
H$\alpha$ emission line luminosities of the BCGs. $c$ means the H$\alpha$ luminosity comes from
\citet{Craw99}, otherwise it is from SDSS DR4. Column 5 is the cooling time of the gas calculated by
R08. Column 6 lists the $u-r$ colour gradient given by R08, while Column 7 is  our own measurement of
the same quantity using imaging data from SDSS} \label{T:X-ray BCG}
\end{center}
\end{table*}

\begin{table*}
\begin{center}
\renewcommand{\arraystretch}{1.}
\begin{tabular}{c c c c c c c c}
\hline \hline
 \multicolumn{1}{c}{} & \multicolumn{1}{c}{M$_r$} & \multicolumn{1}{c}{(g--r)$_{total}$} &
 \multicolumn{1}{c}{(g--r)$_{\mathrm in}$} &
\multicolumn{1}{c}{(u--r)$_{total}$} & \multicolumn{1}{c}{(u--r)$_{\mathrm in}$} &
\multicolumn{1}{c}{(NUV--r)$_{total}$} & \multicolumn{1}{c}{(NUV--r)$_{\mathrm in}$}\\
\hline
     X-ray vs optical&    -0.26& -0.32& -0.66& -0.58&  -0.21& -1.55&  \textbf{-1.39}   \\
 cool-core vs optical&    -0.60& -1.49& -1.26& -0.22&  -0.94& -3.69&  \textbf{-3.51}  \\
\hline \hline
\end{tabular}
\caption{K-S test values (the logarithm of the K-S probability that the two distributions are drawn from an identical parent population) for differences between optically selected BCGs, X-ray selected BCGs and
cool-core X-ray BCGs. It shows that the difference of $(NUV-r)_{in}$ between X-ray selected (especially cool-core) and optical selected BCGs is strong.} \label{T:KStest}
\end{center}
\end{table*}

\subsection{ Consistency with previous work}
\label{subsec:rst3} In the previous subsection, we showed that cool-core and non-cool-core clusters
 segregated rather cleanly in colour space, implying that the presence of young stars in the BCG
is related to the cooling time of the gas in cluster cores.

There have been a number of studies of  X-ray selected clusters
that have  found that
there is a central entropy (or t$_{cool}$) threshold below which
BCGs have significant H$\alpha$ line emission and blue cores
\citep{Bild07,Raff08,Pip08,Cava08}.

We now check whether we reproduce these results using the sample of X-ray BCGs from R08. For this
analysis, we do not impose a cut on the sample at $z\leq$0.1 --  the reason is that this would reduce
the sample size by too much. Table~\ref{T:X-ray BCG} shows the values of the $u-r$  colour gradients
given in R08 and our measurements from the SDSS images. The measurements are generally
consistent given the uncertainties. When available, we take the H$\alpha$ fluxes from \citet{Craw99}
(to be consistent with \citet{Cava08}), otherwise, we adopt the  H$\alpha$ fluxes in the SDSS DR7
emission line catalog released by the MPA/JHU group.

\begin{figure}
\bc  \hspace{-0.8cm} \resizebox{7.5cm}{!}{\includegraphics{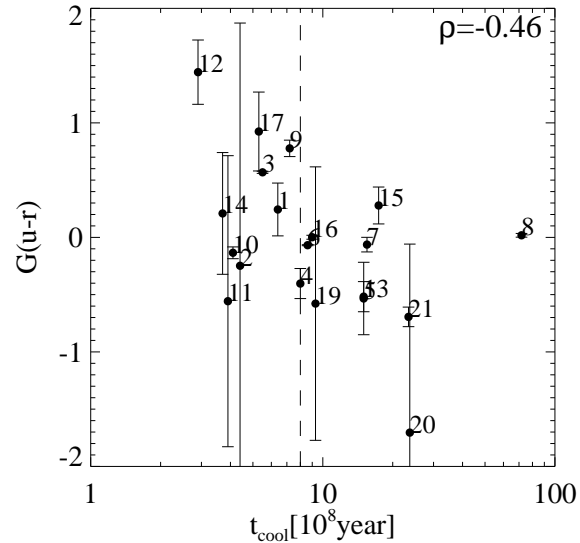}} \caption{ The $u-r$ colour
gradient as defined in R08 (bottom, see text) is plotted as a function of the cooling time of the gas
in units of $10^8$ years. The dashed line marks 0.8 Gyr, which R08 claim  to be the threshold for
finding blue cores. Numbers are used to mark individual BCGs as listed in Table 1.  $\rho$ is
the cross correlation parameter, as defined in Section 4.3.
.} \label{fig:raf1} \ec
\end{figure}

\begin{figure}
\bc  \hspace{-0.8cm} \resizebox{7.5cm}{!}{\includegraphics{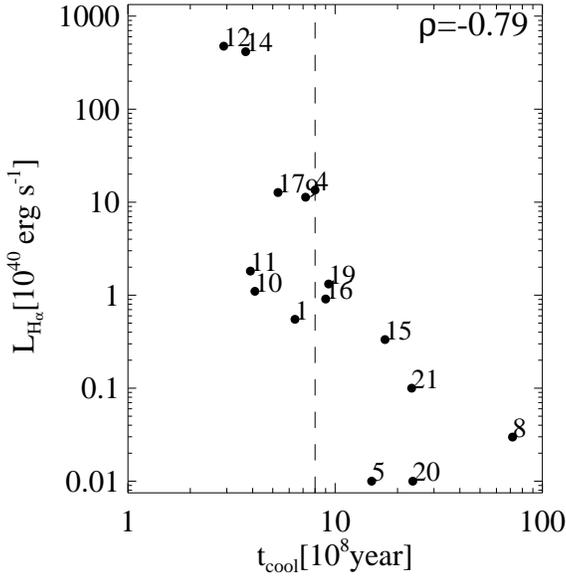}} \caption{ $H\alpha$ line
luminosity is plotted against $t_{cool}$ for the R08 BCG sample. Numbers are used to mark individual
BCGs. $\rho$ is the cross correlation parameter.} \label{fig:raf11} \ec
\end{figure}

\begin{figure}
\bc  \hspace{-0.8cm} \resizebox{7.5cm}{!}{\includegraphics{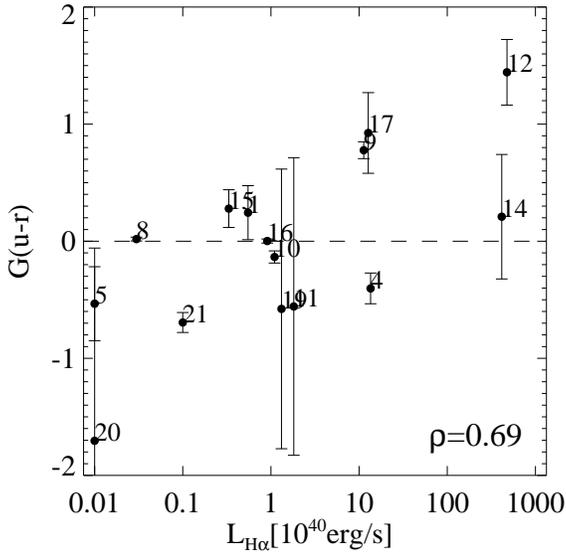}} \caption{The correlation between
$G(u-r)$ and $H\alpha$ line luminosities for the R08 sample. Numbers are used to mark individual BCGs.
$\rho$ is the cross correlation parameter.} \label{fig:raf12} \ec
\end{figure}

We plot  $u-r$ colour and H$\alpha$ luminosity versus t$_{cool}$  in Figures~\ref{fig:raf1} and
~\ref{fig:raf11} respectively. We use the cross correlation parameter $\rho=\langle(X-\bar{X})(Y-\bar{Y})\rangle$ to evaluate the significance of corrleation between parameter $X$ and $Y$. As can be seen, we do verify a threshold $t_{cool}$ value of $\sim$1 Gyr
below which {\em some} BCGs exhibit significant H$\alpha$ luminosity and blue cores. It should be
noted, however, that {\em not all} BCGs with $t_{cool} < 1$ Gyr are strongly star-forming.  There is a
significant scatter in H$\alpha$ luminosity  and colour gradient values and only weak correlation with
the actual value of $t_{cool}$ below the threshhold value.
If we plot $u-r$ colour gradient
versus H$\alpha$ luminosity for these X-ray BCGs (Figure~\ref{fig:raf12}), 
there is a significant 
correlation  between the two quantities (albeit with large scatter). 
We also find similar
correlations between 
$u-r$ colour differences (following the definition from R08) and both
H$\alpha$ luminosity and 
H$\alpha$ equivalent width.

We note that the scatter may arise because 
H$\alpha$ line emission is produced by ionizing
photons from massive stars with lifetimes 
of $\sim 10^7$ years, whereas the $u-r$ colour gradients will
be sensitive to star formation that has taken place 
over Gyr timescales. In addition, it is possible
that part of the H$\alpha$ line emission does not 
originate from HII regions, but is excited by a
central AGN.

\begin{figure}
\bc \hspace{-0.8cm} \resizebox{7.5cm}{!}{\includegraphics{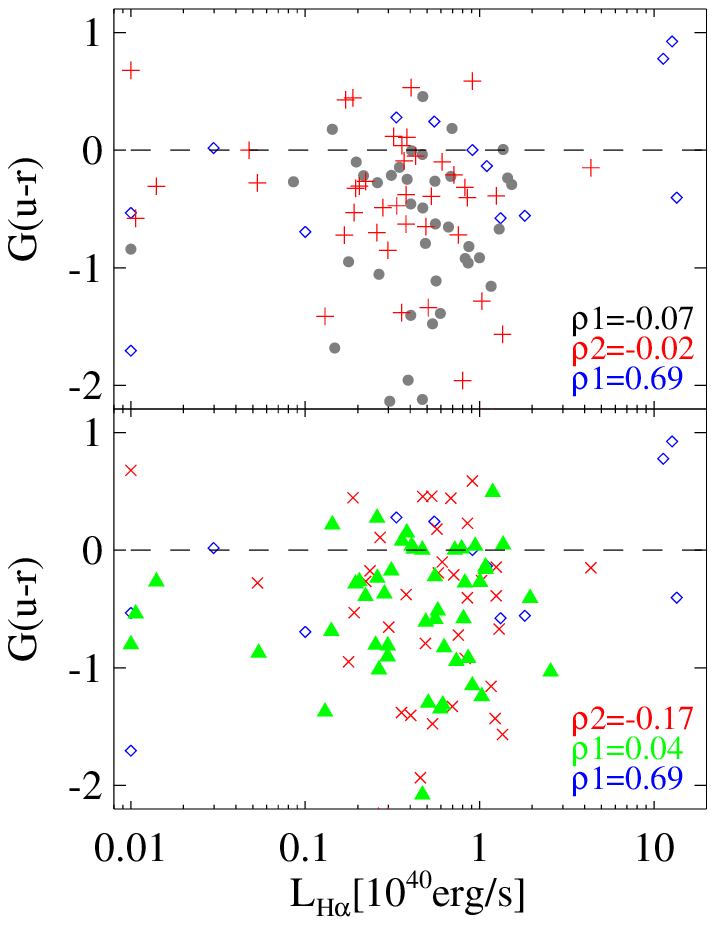}} \caption{The correlation between
$G(u-r)$ and $H\alpha$ line luminosities for the optically selected BCG of vdL07. In the bottom  panel,
the BCGs with $NUV-r<$5.8 are in green (filled triangles), and others are in red (crosses). In the top
panel, the BCGs with ROSAT X-ray luminosities $>10^{43.5}$ erg s$^{-1}$ are in red (pluses), otherwise they are in grey (solid points). Blue
points (diamonds) indicate R08 BCGs. $\rho$ is the cross correlation parameter for the symbols of corresponding colour.} \label{fig:raf2} \ec
\end{figure}

We have also checked whether there is a
correlation between $u-r$ colour gradient 
and H$\alpha$ luminosity in the ROSAT (top) and GALEX MIS (bottom) detected sample of optically-selected
BCGs from vdL07 and  
the results are shown in 
Figure~\ref{fig:raf2}; the lower redshift range of
$z<0.1$ ensures that the $u-r$ colour gradients can be measured
relatively accurately. In the top panel, the points are colour-coded
according to the total X-ray luminosity of the cluster 
(clusters with ROSAT X-ray luminosities greater
than $10^{43.5}$ erg s$^{-1}$ are plotted in red). 
In the bottom panel, the points 
are coded according to the $NUV-r$ colour of
the BCG. We fail to find any correlation between 
colour gradients and H$\alpha$ luminosity for
any of the subsamples defined from the
optically-selected clusters.  

This may not be too surprising in view of the fact
that the optically selected BCGs span a much smaller range
in H$\alpha$ luminosity.
(most L$_{H\alpha}<$1.1$\times10^{40}$ erg/s).
Much of the correlation seen in Figure~\ref{fig:raf12} is driven
by the BCGs with the very higher H$\alpha$ luminosities.
Over the range in H$\alpha$ lumnosity where the R08 sample overlaps
the optically selected clusters, the results are actually in fairly good
agreement.

\section{Summary and Discussion}
\label{sec:discussion}

The main aim of this paper is to investigate the claim that the bright galaxy
located at the center of a massive dark matter halo (BCG) forms stars at a higher
rate than galaxies of similar mass and structure that are not located
at cluster or group centres. Studies of BCGs from X-ray selected cluster
samples have generally supported this claim.  Anlayses  of BCGs from
optically-selected cluster samples have failed to find any significant
enhancement in star formation when compared to control samples of non-BCGs.

Recent work on optically-selected BCGs relied on spectroscopic indicators derived from Sloan Digital
Sky Survey spectra. This has the disadvantage of only  probing the most recent star formation that is
occurring in the cores of the BCGs. In this paper, we analyzed a sample of 113 BCGs with $z< 0.1$  with
optical imaging from SDSS and UV data from GALEX. We convolved the SDSS images to match the resolution
of the GALEX data and measured UV/optical colours both in the inner and outer regions of the BCGs. We
did the same for a smaller sample of X-ray selected BCGs with gas cooling times derived from Chandra
data.

Our main conclusions concerning our sample of optically-selected
BCGs are the following:
\begin {enumerate}
\item We confirm the conclusion of vdL07 that optically-selected BCGs have
similar median $g-r$ and $u-r$ colours as control samples of non-BCGs with the same stellar mass and
concentration. This conclusion holds for colours measured in both the inner and the outer regions of
the galaxies.
\item Optically-selected BCGs have smaller scatter in their $g-r$ and $u-r$ colours
than the control non-BCGs.
\item  Optically-selected BCGs have slighly {\em redder} $NUV-r$ colours in their
centres than the control galaxies.
\item  Correlations between $G(u-r)$ and $H\alpha$ luminosity
are very weak.
\end {enumerate}
Taken together, the three findings lead us to conclude that
optically-selected BCGs are slightly older and have had less recent star
formation than the galaxies in the control sample.

Our main conclusions concerning our sample of X-ray
BCGs are the following:
\begin {enumerate}
\item The optical ($g-r$ and $u-r$) colours of the BCGs in the X-ray
selected cluster sample span a similar range as those in the subset of optical clusters with high
cluster ($\log \sigma > 2.8$) velocity dispersions. Their $NUV-r$ colours, on the other hand, are
significantly bluer.
\item Among the X-ray selected BCGs, those that are located in the centre
of a cool-core cluster (defined to have central cooling times of less than 1 Gyr)
always have bluer optical and UV/optical colours than those that are
located in clusters where the central gas cooling times are long.
\item We confirm that there appears to be a threshold value of the central cooling time
below which star formation does occur in a subset of clusters.
\end {enumerate}

Our main conclusion, therefore, is that the location of a galaxy at the center
of a dark matter halo is not sufficient to determine whether or not it
is currently forming stars -- one must also have information about the
thermo-dynamic state of the gas in the core of the dark matter halo.
These results agree with other similar studies.
\citet{Egami06} found that the majority of the
BCGs are not particularly infrared-luminous compared with
other massive early-type galaxies.
\citet{Edw07} found their red BCGs selected from the SDSS sample
had a low fraction  with emission lines
($11\pm2$ percent), comparable to that found in their control sample
However,  both papers noted that for BCGs
in clusters where cooling times are very short and
the predicted cooling flow rates are high, star formation
is prominently enhanced.

Perhaps the most important lesson to take away from this analysis, is that
BCGs selected from optical cluster surveys and BCGs selected from X-ray
cluster surveys will be different. Low redshift spectroscopic surveys
such as the SDSS are designed to provide a {\em complete} census of
the most massive galaxies out to redshifts of $0.2-0.3$. The fact
that only a relatively small percentage of BCGs in these
surveys have signs of recent star formation must indicate that
the majority of nearby  BCGs do not sit at the centers of dark matter halos in which
central cooling time of the gas is less than 1 Gyr.
As shown by \citet{Best07}, a large fraction (20-30\%) of optically-selected
BCGs host radio-loud AGN, which may play a key role in preventing the gas from cooling.

It remains to be seen whether the same conclusion will hold up at higher redshifts. Many of the most
strongly star-forming BCGs in the current X-ray samples are located at relatively  high redshifts. We
checked the full  R08 sample  and found that the ratio of blue core BCGs to those with negative
gradients increases with redshift: 19:27 for the whole sample, 13:8 for BCGs at $z>$0.1, and 6:3 for
BCGs at $z>$0.2. In addition, \citet{Craw99} found that the fraction  of emission line BCGs increases
with redshift. We note that the cluster sample of vdL07 is defined to lie at  redshifts below  0.1, so
this may explain why the BCGs in this sample are predominantly inactive. On the other hand, only the
most X-ray luminous clusters can currently be detected at higher redshifts and these are likely to be
biased to systems with  short central cooling times. A new generation of both optical spectroscopic
surveys (e.g. SDSS-III's Baryon Oscillation Spectroscopic Survey ($BOSS$)) and deeper X-ray imaging
surveys over large areas of the sky (e.g. the extended R\"{o}entgen Survey with an Imaging Telescope
Array (eROSITA) will be needed before these issues can be fully understood.

\section*{Acknowledgements}
We would like to thank Shiyin Shen who kindly provide us with helpful discussion and ROSAT
cross-matched catalog for our optical BCG sample. We also thank Tim Heckman,  Qi Guo and Cheng Li for
their helpful comments and discussions. We are grateful to the anonymous referee for
a very thorough and insightful review of our manuscript. XK is supported by the National Natural 
Science Foundation of China (NSFC, Nos. 10633020, and 10873012), the Knowledge Innovation Program of 
the Chinese Academy of Sciences (No. KJCX2-YW-T05), and National Basic Research Program of China (973 
Program; No. 2007CB815404).

GALEX (Galaxy Evolution Explorer) is a NASA Small Explorer, launched in April 2003, developed in
cooperation with the Centre National d'Etudes Spatiales of France and the Korean Ministry of Science
and Technology.

Funding for the SDSS and SDSS-II has been provided by the Alfred P. Sloan Foundation, the Participating
Institutions, the National Science Foundation, the U.S. Department of Energy, the National Aeronautics
and Space Administration, the Japanese Monbukagakusho, the Max Planck Society, and the Higher Education
Funding Council for England. The SDSS Web Site is http://www.sdss.org/.

The SDSS is managed by the Astrophysical Research Consortium for the Participating Institutions. The
Participating Institutions are the American Museum of Natural History, Astrophysical Institute Potsdam,
University of Basel, University of Cambridge, Case Western Reserve University, University of Chicago,
Drexel University, Fermilab, the Institute for Advanced Study, the Japan Participation Group, Johns
Hopkins University, the Joint Institute for Nuclear Astrophysics, the Kavli Institute for Particle
Astrophysics and Cosmology, the Korean Scientist Group, the Chinese Academy of Sciences (LAMOST), Los
Alamos National Laboratory, the Max-Planck-Institute for Astronomy (MPIA), the Max-Planck-Institute for
Astrophysics (MPA), New Mexico State University, Ohio State University, University of Pittsburgh,
University of Portsmouth, Princeton University, the United States Naval Observatory, and the University
of Washington.
\label{lastpage}

\bibliographystyle{mn2e}

\end{document}